\documentclass[12pt,a4paper]{article}

\usepackage[scale={.75,.8}]{geometry} 
\usepackage{pstricks}
\usepackage[dvips]{graphicx}
\usepackage{axodraw}
\usepackage[latin2]{inputenc}
\usepackage{epic}
\usepackage{latexsym}
\usepackage{epsf}
\usepackage{epsfig}
\usepackage{amssymb,amsmath}
\usepackage{cite} 
\usepackage{color}

\newcommand{\fref}[1]{fig.\ \ref{f.#1}} 
\newcommand{\eref}[1]{eq.\ (\ref{e.#1})} 
\newcommand{\erefn}[1]{ (\ref{e.#1})}
 
\newcommand{\aref}[1]{\ref{a.#1}}
\newcommand{\sref}[1]{Section \ref{s.#1}}
\newcommand{\cref}[1]{Chapter \ref{c.#1}}

\def\nn{\nonumber \\}  
\newcommand{\nl}{& \nonumber \\ &}

\def\ds{\displaystyle}

\def\beq{\begin{equation}} 
\def\eeq{\end{equation}} 
\def\bea{\begin{eqnarray}}  
\def\eea{\end{eqnarray}}  
\newcommand{\bal}{\begin{align}}
\newcommand{\eal}{\end{align}}   
\def\ba{\begin{array}}  
\def\ea{\end{array}}   
\def\bi{\begin{itemize}}  
\def\ei{\end{itemize}}  
\def\ben{\begin{enumerate}}  
\def\een{\end{enumerate}}  
\def\beq{\begin{equation}}  
\def\eeq{\end{equation}}  
\def\bc{\begin{center}}
\def\ec{\end{center}} 
 \def\bt{\begin{table}}
\def\et{\end{table}}  
 \def\btb{\begin{tabular}}
\def\etb{\end{tabular}}  
\newcommand{\refeq}[1]{\mbox{(\ref{#1})}}

\def\co{{\mathcal O}}

\def\mass2{mass${}^2$}

\def\ads{{\mathrm A \mathrm d \mathrm S}}

\def\ra{\rangle}
\def\la{\langle}  

\def\pa{\partial}

\def\hc{{\rm h.c.}}

\def\eps{\epsilon}

\makeatletter
\def\secteqno{\@addtoreset{equation}{section}%
\def\theequation{\thesection.\arabic{equation}}}

\begin{document}

\secteqno

\pagestyle{empty}
\begin{flushright}
CERN-PH-TH/2006-220\\ 

{\bf \today}
\end{flushright}
\vspace*{5mm}
\begin{center}

\renewcommand{\thefootnote}{\fnsymbol{footnote}}

{\large {\bf Holography, Pad\'e Approximants and Deconstruction 
}} \\ 
\vspace*{1cm}
{\bf Adam~Falkowski}$^{\rm a),b)}$\footnote{Email:adam.falkowski@cern.ch}
and
{\bf Manuel~P\'erez-Victoria}$^{\rm a)}$\footnote{Email:mpv@cern.ch}

\vspace{0.5cm}

a) CERN Theory Division, CH-1211 Geneva 23, Switzerland \\
b) Institute of Theoretical Physics, Warsaw University, \\ Ho\.za 69,
00-681 Warsaw, Poland

\vspace*{1.7cm}
{\bf Abstract}
\end{center}
\vspace*{5mm}
\noindent

{
We investigate the relation 
between holographic calculations in 5D and 
the Migdal approach to 
correlation functions in large-$N_c$ theories.
The latter employs Pad\'e approximation to  extrapolate
short-distance correlation functions to  
large distances. 
We make the Migdal/5D relation more precise by quantifying the correspondence between  Pad\'e
approximation and  the background and boundary conditions in 5D. 
We also establish a connection between the Migdal approach and the models of deconstructed
dimensions.  
}
\vspace*{1.0cm}
\date{\today}


\vspace*{0.2cm}
 
\vfill\eject
\newpage

\setcounter{page}{1}
\pagestyle{plain}

\renewcommand{\thefootnote}{\arabic{footnote}}
\setcounter{footnote}{0}

\section{Introduction}

Holography relates strongly-coupled gauge theories to weakly-coupled
theories in higher dimensions.  
The original conjecture~\cite{Maldacena} connects type IIB string
theory in the gravitational $\ads_5 \times S_5$ background to the 4D
$\mathcal{N}= 4$ $SU(N_c)$ superconformal field theory. This
correspondence can be extended to other
asymptotically-AdS~\cite{adscft} spaces, 
and examples of geometries in which conformal invariance is broken in the IR
are known (see, for instance,~\cite{nonconformaladscft} and references
therein). 
From a phenomenological point of view, it is often sufficient to
consider the simpler relation between non-supersymmetric 5D field 
theories in a Randall--Sundrum \cite{rasu} background and 4D
large-$N_c$ strongly-coupled theories with  conformal invariance
spontaneously broken 
\cite{adscftrs}.     
This can be applied, for example, in studying properties
of QCD at large $N_c$ \cite{adsqcd} or to clarify the physics of
electroweak symmetry breaking by strong dynamics~\cite{higgs}.   

The 5D setup involves a slice of $\ads_5$ truncated by two branes at
$z = z_{\mathrm{UV}}$ and $z=z_{\mathrm{IR}}$.    
The bulk hosts 5D gauge and, possibly, other spin fields, which
represent composite operators of the dual CFT.    
One way to probe the dynamics of such theories is to compute UV
boundary correlators of the 5D bulk fields.  
For example, the 1PI two-point correlation function of boundary gauge
fields is given by the expression  
\beq
\label{e.holoint}
\int d^4x e^{- i p x} \la A_\mu(x) A_\mu(0) \ra_{\mathrm{1PI}} = 
\left (-\eta_{\mu\nu}+ {p_{\mu}p_{\nu} \over p^2}  \right )\Pi (p^2), 
\qquad  
\Pi (p) \sim \left. \frac{\partial_z G(z,p)}{G(z,p)}\right|_{z = z_{\mathrm{UV}}} \, ,
\eeq  
where $G(z,p)$ is a  solution of the 5D equations of motion subject to
appropriate boundary conditions at the IR brane.   
The holographic dictionary relates this boundary correlator to the
connected two-point correlation function of a conserved global
symmetry current of the 4D CFT.   
The poles of this correlation function are interpreted as resonances
of the CFT.    

There exists another, seemingly unrelated approach to computing
correlation functions in 4D strongly coupled large-$N_c$ theories,
which was proposed long ago by Migdal~\cite{Migdal}.  
The program of Migdal aims at reproducing the gauge theory correlators at low $p^2$ using information about the deep Euclidean regime. 
The main input is the non-analytic behaviour of the correlation functions at large Euclidean momenta,  where the correlators, at leading order, exhibit a conformal behaviour, 
$\lim_{p^2 \to -\infty} \Pi(p) \equiv f(p^2) \sim -p^{2n} \log (-p^2)$. 
This asymptotic expression is approximated by a ratio of two
polynomials of degree $N$,  
$f(p^2) \approx R_N(p^2)/S_N(p^2)$, by means of a Pad\'e approximation.
Finally, the result is analytically continued to low time-like $p^2$
and the large-$N$ limit is taken. 

It was recently pointed out by Shifman~\cite{shifman} and
Voloshin, and shown in detail by Erlich et al.~\cite{erkrlo}, that Migdal's
procedure gives results similar to those of 5D holographic computations.  
Indeed, performing a series of Pad\'e approximations on the input
function $f(p^2) = -\log(-p^2/\mu^2)$, one obtains \cite{erkrlo}, for
large $N$,  
\beq
f(p^2) \to  
{-\log (p^2/\mu^2) J_0 (2 N p/\mu) + \pi Y_0 (2 N p /\mu) 
\over  
J_0 (2 N p/\mu)} 
\eeq 
The same result would be  obtained if we performed a computation of
the boundary gauge field correlator in the Randall--Sundrum
spacetime, with $z_{\mathrm{IR}} = 2 N/\mu$ and $z_{\mathrm{UV}}=1/\mu \to 0$. 
In fact, the similarity of the two approaches is not restricted to $\ads_5$. 
For example, Pad\'e approximations of the input function $f(p^2) = (-p^2/\mu^2)^{-1/2}$ lead to a result that again coincides with the 5D boundary gauge correlator, but this time computed in the 5D Minkowski background.\footnote{%
This was first observed by Gherghetta, Pomarol and Rattazzi~\cite{pc}.
}   

This coincidence seems quite mysterious, as the Migdal approach never
makes any reference to extra dimensions. Of course, the main
ingredient of both methods, and the one that allows us to make
contact with a large-$N_c$ theory, is that the correlation functions
are meromorphic. But this does not determine them uniquely. 
Actually, the success of the Migdal approach and its relation
to holographic calculations raises a number of questions: 
\ben 
\item Why, in the first place, can the series of Pad\'e approximants
  be interpreted as physical correlation functions of a large-$N_c$
  gauge theory? 
The physical interpretation of the results 
obtained by Migdal is possible because the Pad\'e approximants have
  only simple poles, 
  with negative residues on the positive real $p^2$ axis.   
This is not a generic feature of Pad\'e approximants. 
\item 
Why, in the large-$N$ limit, do we recover correlators characteristic
of theories in 5D?   
The distinguishing  feature of 5D models is locality in the fifth coordinate.  
How is this locality encoded in the Migdal approach?  
\item Which 5D geometries and which IR boundary conditions can be
  reproduced by the Migdal approach?   
\item Does the Migdal procedure at finite $N$ also correspond to some
  physical setup?  
There is a tantalizing similarity between the correlators computed in
the Migdal approach and those obtained in
  deconstruction~\cite{dec}. Both express 
the polarization function as a ratio of two polynomials in $p^2$, which
converges to a non-analytic function for large $N$ and large Euclidean
  momentum (see \cite{blfape}
for a computation in deconstructed $\ads_5$).  
Is there a precise quantitative connection between the two approaches?  
\een 

In this paper we provide answers to these questions. 
The first one was, 
in fact, already addressed in Migdal's  original papers~\cite{Migdal}.
It turns out that the nice physical properties of Pad\'e
approximants arise because the input function belongs to the class of
so-called Stieltjes functions.
In the mathematical literature, Pad\'e approximants of Stieltjes functions
have been extensively studied, and a connection to orthogonal
polynomials has been established.  
The physical properties of  Pad\'e approximants are 
intimately related to certain familiar properties of  orthogonal
polynomials. 

One well-known property of orthogonal polynomials is that they
satisfy a  second-order recursion relation in the polynomial degree:  
$\pi_{N+1}(w)=(a_n w +b_n)\pi_N(w) - c_n \pi_{N-1}(w)$. 
We show that the connection between Pad\'e approximation and
orthogonal polynomials implies that the polynomials entering the
Pad\'e approximants also satisfy a second-order recursion relation:
\beq 
T_{N+1}(p^2)= A_N(p^2) T_N(p^2) - B_N(p^2)T_{N-1}(p^2) \, ,
\eeq 
where $T= R,S$. Hence, Pad\'e approximants implement automatically
a form of locality in the discrete space of the polynomial degrees. 
For large $N$, this recursion relation  reduces to a second order
differential equation, whose form is analogous to equations of motion
in 5D theories.  
We discuss the necessary condition for the large-$N$ limit of the
Migdal approach to correspond to sensible 5D setups and find which
geometries and which boundary conditions can be matched. 
We shall see that the precise manner in which the limit is taken is important.

Finally, we study the relation between Migdal's approach at finite
$N$ and deconstruction.  
Deconstruction is a four-dimensional framework that involves a product 
gauge group $G^N$ and a set of bifundamental non-linear sigma model
fields (the links) \cite{dec}. 
Deconstruction  models are parametrized by a set of gauge couplings
$g_j$ and decay constants $v_j$.  
For large $N$ such  setup is related to a 5D gauge theory with the
gauge group $G$, where the fifth dimension is latticized.  
Each choice of $g_j$ and $v_j$ on the  deconstruction side corresponds
to some 5D warped geometry,  and the dictionary between the two
frameworks has been established \cite{faki}.  
In this paper we quantify the relation between deconstruction and the
Migdal approach. 
We show that, given the coefficient of the recursion relation for
Pad\'e approximants, we can identify the deconstruction parameters
that yield the same polarization function as the Migdal approach.  
An unexpected result is that deconstruction models  directly related
to the Migdal approach are non-minimal; they must include the kinetic
mixing between neighbouring gauge fields from the product group. This,
however, corresponds to an irrelevant operator in the continuum limit.

The paper is organized as follows. 
In \sref{mae} we present several examples that illustrate the
connection of the Migdal approach with 5D theories.   
In \sref{paop} we review the mathematical results connecting Pad\'e
approximants of Stieltjes functions to orthogonal polynomials.  
The large-$N$ limit of Pad\'e approximants of Stieltjes functions is
studied in \sref{ml} and the relation to 5D theories is quantified in
\sref{mc}.  
In \sref{dh} we discuss the connection of Migdal approximation at finite
$N$ to 4D deconstruction models.  
Section \ref{s.c} contains our conclusions, and in the appendix we
derive the holographic formula for two-point functions in
deconstruction. 

\section{Migdal's approximation in examples}
\label{s.mae}

For two-point correlation functions, the deep Euclidean limit $p^2 \to
-\infty$ is a function $f(t)$ depending on a variable $t=p^2/\mu^2$,
where  
$\mu$ is an arbitrary renormalization scale. 
The function $f(t)$  has a branch cut along
the positive real axis and contains a perturbative piece plus power
corrections induced by the condensates. We will be interested in the
perturbative part only. 
Migdal proposes to compute Pad\'e approximants of $f(t)$ at some
Euclidean point $-\lambda < 0$. 
The Pad\'e approximant is given by a ratio of two polynomials of degrees $M$ and
$N$, such that its Taylor expansion around $t = -\lambda$ matches that of $f(t)$ to order $(t+\lambda)^{M+N+1}$. 
Under certain assumptions, this series of approximants converges to $f(t)$ when 
$M,N \rightarrow \infty$.  
The idea of Migdal is to take instead a combined limit 
$N\rightarrow \infty$, $t \rightarrow 0$, keeping 
$\tilde{t}=t N^2$  and $M-N$ fixed. 
For finite momentum $p$, this amounts to sending $N$ and $\mu$
to infinity with 
$\tilde{\mu}=\mu/N$ fixed. 
We call this {\em the Migdal limit}. 
In this limit, the spacing between the poles of the Pad\'e approximants
is controlled by the scale $\tilde{\mu}$ introduced in the limiting
process. Intuitively, even though the spacing between poles goes to zero
at large $N$, so as to reproduce the branch cut, zooming in the small
$t$ region simultaneously allows to resolve them. 
It turns out that, for the functions $f(t)$ of interest, the limiting
expression has only simple poles 
located at timelike momenta and with negative residues. 
Therefore, the Migdal limit of $f(t)$ can be interpreted as the
 complete two-point function in the 
large~$N_c$ limit.
By construction, this function has the correct deep Euclidean behaviour. 

In this section we present several examples of the Migdal procedure. 
The details of the calculations are postponed until \sref{paop}, where
we present a generalized approach to this kind of computation. 

We start with the example discussed in~\cite{erkrlo}.
The two-point correlation function of two conserved vector currents has the general form 
\beq
\Pi_{\mu\nu}(p) =  \left(\frac{p_\mu p_\nu}{p^2}-
\eta_{\mu\nu}\right)\, \Pi(p^2) . 
\eeq
For the vector and axial currents in QCD,  
the leading perturbative contribution to $\Pi(p^2)$ at
large Euclidean momentum $p^2<0$ is proportional to 
$-p^2\log(-p^2/\mu^2)$.\footnote{We take the principal branch of the logarithm
with the branch cut along the real negative axis (argument
$\theta\in(-\pi,\pi]$), and define accordingly non-integer powers. The
signature of the metric is mostly minus. Physical amplitudes at
time-like momenta are evaluated with $p^2$ right above the positive real axis.}
Ignoring  multiplicative constants, we take $f(t+1)=-\log(-t)$. 
Next, we approximate $f(t+1)$ by $\Pi_N=t [N/N]_f$, where $[N/N]_f=R_N/S_N$ is the Pad\'e approximant to
$f(t+1)$ at $t=-1$ and  $R_N,S_N$ are polynomials in $t$ of degree $N$. 
These polynomial can be determined to be 
\beq
\label{e_logpade1}
R_N(t)=  (t+1)^N \bar{P}_N\left(\frac{1-t}{1+t}\right), \qquad
S_N(t) = (t+1)^N P_N\left(\frac{1-t}{1+t}\right)  \, ,
\eeq
with $P_N$ the Legendre polynomial of degree $N$ and $\bar{P}_N$ the associated Legendre polynomial (see the next section for the definition of associated orthogonal polynomials). 
The factors $(t+1)^N$ in~\refeq{e_logpade1}
cancel out in the quotient; their role is only to make $R_N$ and $S_N$
polynomials in $t$.  
In the Migdal limit, $R_N$ and $S_N$ approach  
\begin{align}
\label{e_loglimit}
& R(t)=-\log (t) J_0 \left(2 \sqrt{\tilde{t}}\right) + \pi Y_0 \left(2
\sqrt{\tilde{t}}\right) , \\  
& S(t) = J_0\left(2 \sqrt{\tilde{t}}\right) .
\end{align}
On the other hand, a calculation of the two-point boundary correlator
for a gauge field in $\ads_5$ with Neumann boundary conditions  at the
IR brane yields \cite{adscftrs}, in the limit $z_{\mathrm{UV}} \to 0$,   
\beq
\Pi(p^2) \sim p^2 
{-\log (p z_{\mathrm{UV}}) J_0 \left(p z_{\mathrm{IR}}\right) 
+ \pi Y_0 \left(p z_{\mathrm{IR}}\right) \over   
 J_0\left(2 p z_{\mathrm{IR}}\right) } 
\eeq
We see that identifying  
$\mu \leftrightarrow z_{\mathrm{UV}}^{-1}$, $\tilde{\mu} \leftrightarrow  2
z_{\mathrm{IR}}^{-1}$  
the result $\Pi_{\mathrm{Migdal}}=t R/S$ agrees precisely with the 
corresponding holographic calculation. 
Note that, according to the dictionary above and the definition
of $\tilde{\mu}$, $2 N$ corresponds to the inverse warp factor
$z_{\mathrm{IR}}/z_{\mathrm{UV}}$. 

At this point it is natural to wonder why the Pad\'e approximant
chooses Neumann conditions. 
Our next example shows that we have actually made this choice when
identifying the function $f(t)$, leaving the factor $t$ outside of the
Pad\'e approximation.    
Indeed, let us take instead $f(t+1)=-t \log (-t)$, compute the $(N+1,N)$
Pad\'e approximant to $f(t+1)$ at $t=-1$ and define $\Pi_N(t)=
[N+1/N]_f=R_{N}/S_N$ . It is clear that this keeps the same
asymptotic 
function as before. The reason for increasing the degree of the
numerator is to improve the convergence at large $|t|$. This technical
point will be clarified in the next section. The result is 
\begin{align}
\label{e_tlogpade}
& R_N(t)=(t+1)^{N+1} \left[-P^{(1,0)}_N \left(\frac{1-t}{1+t}\right) +
  \bar{P}^{(1,0)}_N \left(\frac{1-t}{1+t} \right) \right] , \\ 
& S_N(t) = (t+1)^N P^{(1,0)}_N\left(\frac{1-t}{1+t}\right) ,
\end{align}
with ($\bar{P}^{(\alpha,\beta)}_N)$ $P^{(\alpha,\beta)}_N$
(associated) Jacobi polynomials. 
The Migdal limit yields, up to normalization, 
\begin{align}
\label{e_tloglimit}
& R(t)= -t \log (t) \frac{1}{\sqrt{\tilde{t}}} J_1 \left(2
\sqrt{\tilde{t}}\right) + \pi t Y_1 \frac{1}{\sqrt{\tilde{t}}} \left(2
\sqrt{\tilde{t}}\right) , \\ 
& S(t) = \frac{1}{\sqrt{\tilde{t}}} J_1\left(2 \sqrt{\tilde{t}}\right) .
\end{align}
The result $\Pi_{\mathrm{Migdal}}=R/S$ is identical to the holographic
one for large $z_{\mathrm{UV}}\to 0$, with the same identifications as before, but this
time with Dirichlet boundary conditions at the IR brane. 

Now, let us consider a different asymptotic behaviour in the deep
Euclidean regime.  
We assume that  the polarization  function of two vector currents
approaches $-(-t)^{1/2}$ for  
$t \to \infty$.     
This behaviour is very different from the one encountered in QCD. 
Instead, it is the prediction of a holographic calculation in an
asymptotically flat space.  
Let us  first compute $\Pi_N(t)= t [N/N]_f= t R_N/S_N$ with
$f(t+1)=(-t)^{-1/2}$. We obtain  
\begin{align}
&R_N(t)=(t+1)^N \left[P^{-1/2,1/2}_N\left(\frac{1-t}{1+t}\right) +
    \bar{P}^{-1/2,1/2}_N\left(\frac{1-t}{1+t}\right) \right], \\ 
&S_N(t)=(t+1)^N P^{-1/2,1/2}_N\left(\frac{1-t}{1+t}\right) ,
\end{align}
and in the Migdal limit,
\begin{align}
&R(t)= \frac{1}{\sqrt{t}} \sin \left(2 \sqrt{\tilde{t}}\right), \\
&S(t)= \cos \left(2 \sqrt{\tilde{t}}\right) .
\end{align}
In this case, $\Pi_{\mathrm{Migdal}}=t R/S$ agrees with the
holographic two-point function in 5D Minkowski space with Neumann boundary conditions at the IR brane!
Choosing instead $f(t+1)=-(-t)^{1/2}$ and $\Pi_N(t)=[N/N]_f$, we
arrive at  
\begin{align}
&R_N(t)=(t+1)^N \left[-P^{1/2,-1/2}_N\left(\frac{1-t}{1+t}\right) +
    \bar{P}^{1/2,-1/2}_N\left(\frac{1-t}{1+t}\right) \right], \\  
&S_N(t)=(t+1)^N P^{1/2,-1/2}_N\left(\frac{1-t}{1+t}\right) ,
\end{align}
with the Migdal limit
\begin{align}
&R(t)= -\cos \left(2 \sqrt{\tilde{t}}\right), \\
&S(t)= \frac{1}{\sqrt{t}} \sin \left(2 \sqrt{\tilde{t}}\right) .
\end{align}
As could be already guessed, $\Pi_{\mathrm{Migdal}}=R/S$ is in this case
the same as the holographic two-point function in 5D Minkowski space with
Dirichlet boundary conditions on the IR brane. 

We have given several examples in which the Migdal procedure in a
mysterious way reconstructs the full 5D correlation functions  from
information about the deep Euclidean limit. 
One should be however aware that this ``magic'' does not always work. 
For example, let us try to reproduce the result for the two-point
function of a scalar operator of conformal dimension~3, which is dual 
to a massive scalar in $\ads_5$, with mass $M^2=-3 k^2$. The asymptotic
result of the AdS calculation in the deep Euclidean is, up to a
constant term, $-t \log (-t)$. This is the same as for gauge bosons,
except for the constant term. But the Pad\'e approximant is determined
essentially by the non-analytic piece of the function, as we discuss
in the next section. Therefore, choosing $f(t+1)=-\log (-t)$ 
(and multiplying by $t$ at the end) or $f(t+1)=-t \log(-t)$, we arrive at the
same Pad\'e approximants as in the first two examples. 
In the second case, when a local polynomial term is added, we reproduce the
holographic result for the massive scalar with Dirichlet boundary
conditions. 
However, the first choice does not give the holographic function for
Neumann boundary conditions, but rather one with mixed boundary
conditions on the IR brane.

\section{Pad\'e approximants and orthogonal polynomials}
\label{s.paop}

In this section, following mathematical literature\cite{padebook}, we present a more systematical approach to Pad\'e approximants.
Given a function $f(s)$ with a Taylor expansion at $s=0$  we can define the Pad\'e approximant as follows. 
We introduce two polynomials $R_N^J(s)$, $S_N^J(s)$ of degrees $N+J$, $N$, respectively. 
We choose them such that their ratio has a Taylor expansion at $s=0$ that matches the Taylor expansion of   $f(s)$ up to terms of order $s^{2N+J+1}$: 
\beq
\frac{R_N^J(s)}{S_N^J(s)} = f(s) + O(s^{2N+J+1}) \, 
\label{e_padedef}
\eeq
We also assume $S_N^J(0) \neq 0$. 
The Pad\'e approximant is defined as 
\beq
[N+J/N]_f(s) = \frac{R_N^J(s)}{S_N^J(s)} \, .
\eeq
We will often omit the specification of the function and/or the
variable, and write simply $[N+J/N]$.
Note that $R_N^J$ and $S_N^J$ are determined up to normalization only, and
that both depend on $N$ and $J$. In the
following we will restrict the input function $f$ to be a Stieltjes
function. 
The input functions we studied in the previous section belong to this class
 up to a certain number of subtractions, as we discuss at the end of this section. 
A Stieltjes function is defined by the Stieltjes integral representation
\beq
f(s) = \int_0^{\lambda^{-1}} \frac{d \phi(u)}{1-s u}  \label{e_Stieltjesdef}
\eeq
where $\phi(u)$ is a bounded, nondecreasing function on $0\leq
u<\infty$, with finite real-valued moments 
\beq
f_j=\int_0^\infty u^j d \phi(u) \, , ~ j=0,1,2,\ldots \, ,
\eeq
and $\lambda>0$.
An expansion of \refeq{e_Stieltjesdef} in power series at $s=0$ gives
the Taylor series
\beq
\label{e_series}
f(s)=\sum_{j=0}^\infty f_j s^j,
\eeq
which converges in the open disk $|s|<\lambda$. We
assume in the following that the function $\phi(u)$ is strictly
increasing (and hence the measure strictly positive) except for, at
most, a discrete number of points. Then, $f(s)$ has a branch cut along
$\lambda<s<\infty$.

\begin{figure}[t]
\begin{center}
\scalebox{.5}{%
\fcolorbox{white}{white}{
 \begin{picture}(234,208) (77,-26)
   \SetWidth{0.5}
   \SetColor{Black}
   \DashArrowLine(230,81)(300,81){2}
   \DashArrowArc(227.77,72)(8.79,75.32,284.68){2} 
   \DashCArc(188,72)(110,6,180){2} 
   \Photon(224,72)(301,73){2}{8} 
   \DashArrowLine(300,64)(230,64){2}
   \LongArrow(188,-35)(188,190) 
   \SetWidth{0.6}
   \DashCArc(188,72)(110,-180,-6){2} 
   \SetWidth{0.5}
   \LongArrow(77,72)(309,72) 
   \Text(210,68)[lb]{\small{\Black{$\lambda$}}}
 \end{picture}
}
}
\end{center}
\caption{The contour of integration $\Gamma$.}
\label{fig_contour}
\end{figure}
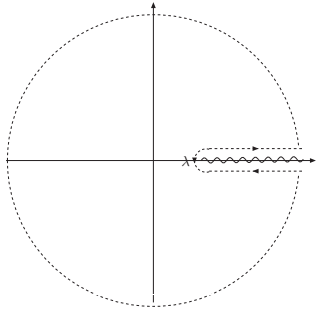

There is a remarkable connection between Pad\'e approximants and
orthogonal polynomials, which we derive next (for reviews of orthogonal polynomials see, for instance, \cite{orthogonalpolynomials}). From the defining
equation~\refeq{e_padedef}, it is clear that  
\beq
\frac{d^m}{ds^m}\left.\left(f(s) S_N^J(s) \right) \right|_{s=0} = 0,
~~ m=N+J+1,\ldots,2N+J \, .  
\eeq
Then we can use Cauchy's integral formula to write these conditions in
the form of contour integrals: 
\beq
\int_\Gamma dz \frac{S_N^J(z) f(z)}{z^{m+1}}=0, ~~  m=N+J+1,\ldots,2N+J \,
\eeq
with the path $\Gamma$ displayed in Fig.~\ref{fig_contour}. The
integral along the small semicircle at $\lambda$ vanishes since, from
its definition~\refeq{e_Stieltjesdef}, $f(z)$ has at most a logarithmic
singularity at $z=\lambda$. The integral along the big circle also
vanishes when $J \geq -1$. Finally, the remaining integrals above and
below the branch cut cancel except for the discontinuity in the
imaginary part of $f(z)$. On the other hand,
from~\refeq{e_Stieltjesdef} this jump is related to the measure
$d\phi(u)=\phi^\prime(u)du$ by 
\beq
\phi^\prime(u^{-1}) = \frac{u}{2\pi i} \big(f(u+i \eps)-f(u-i \eps)
\big) \, , \lambda < u < \infty.   
\eeq
Therefore, for $J\geq -1$ and any $N\geq 0$, 
\beq
\int_\lambda^{\infty} \phi^\prime(u^{-1}) \frac{S_N^J(u)}{u^{m+2}}=0,
~~  m=N+J+1,\ldots,2N+J . 
\eeq
Changing variables to $w=u^{-1}$ and shifting $m\rightarrow m-N-J-1$, we arrive at
\beq
\label{e_orthogonal}
\int_0^\lambda dw W^J(w) w^m \left(w^N S_N^J(w^{-1})\right)=0, ~~  m=0,\ldots,N-1,
\eeq
with $W^J(w)=w^{J+1} \phi^\prime(w)$. The factor in parenthesis is a
polynomial of degree $N$. Eq.~\refeq{e_orthogonal} shows that the set
of polynomials 
\beq
\pi_N^J(w) = w^N S_N^J(w^{-1}), ~~ N=0,1,\ldots
\eeq
is a system of orthogonal polynomials over the interval
$(0,\lambda^{-1})$ with weight $W^J(w)$. We shall also use the
notation $d\phi^J(w)=w^{J+1}d\phi(w)=W^J(w)dw$. Conversely, 
\beq
\label{e.mdopr} 
S_N^J(s) = s^N \pi_N(s^{-1}) \, .
\eeq
This determines the denominators of the Pad\'e approximant $[N+J/N]$
up to normalization. 
To calculate the numerators, let us define the function 
\beq
F^J(w)=\int_0^{\lambda^{-1}} \frac{d\phi^J(u)}{w-u} 
\eeq
and the associated orthogonal polynomials
\beq
\label{e_numdef}
\rho_N^J(w) = \int_0^{\lambda^{-1}} d\phi^J(u) \frac{\pi_N(w)-\pi_N(u)}{w-u} \, ,
\eeq
which have degree $N-1$. Then,
\beq
\pi_N^J(w) F^J(w) = \rho_N^J(w) + \Delta_N^J(w) \, , 
\eeq
where
\begin{align}
\label{e_delta}
\Delta(w) & =  \int_0^{\lambda^{-1}} d\phi^J(u) \frac{\pi_N(u)}{w-u} \nn
& = \frac{1}{w} \int_0^{\lambda^{-1}} d\phi^J(u) \left[1+\frac{u}{w}+\cdots+\left(\frac{u}{w}\right)^{N-1}+\left(\frac{u}{w}\right)^N \left(1-\frac{u}{w}\right)^{-1} \right] \nn
& = w^{-m-1} \int_0^{\lambda^{-1}} d\phi^J(u) \frac{u^N}{1-u/w} \pi_N
\end{align}
is of order $w^{-N-1}$ at large $|w|$. On the other hand, expanding
the integrand in the definition of $F(s)$ at $u=0$ we find  
\beq
\label{e_substraction}
F^J(s^{-1})=s^{-J}\left( f(s) - \sum_{j=0}^J f_j s^j \right)\, .
\eeq 
Therefore, we obtain
\beq
f(s) = \sum_{j=0}^J f_j s^j + \frac{s^J
  \rho_N^J(s^{-1})}{\pi_N^J(s^{-1})} + O(s^{2N+J}) ,\eeq 
so that the numerator of the Pad\'e is 
\beq
R_N^J(s) = S_N^J(s) \sum_{j=0}^J f_j s^j + s^{N+J} \rho_N^J(s^{-1}) . 
\eeq
We see that it is a polynomial of degree $N+J$, as required.

Putting all the pieces together, we have shown that the Pad\'e approximants of a Stieltjes function can be expressed as 
\beq
\label{e_finalpade}
[N+J/N]_f(s)= \sum_{j=0}^J f_j s^j + \frac{s^J
  \rho_N^J(s^{-1})}{\pi_N^J(s^{-1})} .
\eeq 

An important consequence of their relation to orthogonal
polynomials is that the Pad\'e approximants (with $J\geq-1$) of Stieltjes
functions have only simple poles inside the open
interval $(\lambda,\infty)$, and all the residues are negative. This is
the basic property which allows to make contact with large-$N_c$ and
with holography at the classical level. It also follows
from~\refeq{e_delta} that the series of Pad\'e  approximants
$[N+J/N]$ of $f(s)$ converge to $f(s)$ in the limit $N\rightarrow
\infty$, for all $J\geq -1$, in the region $|s|<\lambda$. (This is not
necessarily so for the Pad\'e approximants of a general function.) The
rate of convergence is geometric, as $\Delta(w)\sim 
O\left(w^{-N-1}\right)$. This can be extended, with weaker rate of
convergence, to the domain~$\mathbb{C}\backslash 
[\lambda,\infty)$.


The orthogonal polynomials on the real line obey a three-term
recurrence relation of the form 
\beq
\label{e_recrel}
\pi_{N+1}(w)=(a_N w +b_N)\pi_N - c_N \pi_{n-1} ,
\eeq
with $a_N>0$, $b_N$ real constants and $c_N=(a_N h_N)/(a_{N-1}
h_{N-1})$, where 
\beq
h_N = \la \pi_N| \pi_N \ra = \frac{\int dw W(w) \pi_N(w)^2}{\int dw W(w)}
\eeq 
is the squared norm. We shall call $\bar{a}_N$, $\bar{b}_N$,
$\bar{c}_N=\bar{a}_N/\bar{a}_{N-1}$ the coefficients with normalization
$h_N=1$. The initial conditions for the recurrence relation are 
$\pi_{-1}=0$ (or $c_0=0$) 
and $\pi_0=C$ a nonvanishing constant. From their definition, it is
clear that the associated polynomials $\rho_N$ satisfy the same
recurrence relation but with initial conditions $\rho_0=0$ and
$\rho_1=a_0 C {\int dw W(w)}$. From~\refeq{e_recrel}, a three-term
recurrence relation for Pad\'e numerators and
denominators~\refeq{e_paderecrel} follows: 
\beq
\label{e_paderecrel}
T_{N+1}^J(s)=(a_N^J + b_N^J s) T_N^J(s) - c_N^J s^2 T_{N-1}^J(s) \, ,
\eeq
where $T=R,S$ and we have explicitly indicated the dependence of the
coefficients on $J$. The equation is identical for $S$ and $R$, but
the initial conditions are not. These follow from the ones for the
orthogonal polynomials and associated orthogonal polynomials,
respectively. For instance, when $J=-1$, $R_0^{-1}=0$,
$R_1^{-1}=\mathrm{const}$, $S_{-1}^{-1}=0$,
$S_0^{-1}=\mathrm{const}$. Note the factor $s^2$ in the last term,
which makes the 
equation different from the one for orthogonal polynomials. This extra
factor is very relevant for our purposes, as it has a non-trivial role
in the Migdal limit. It can be shown that, because the weight function
$W$ has a compact support, the combinations $1/\bar{a}_N^J$
and $\bar{b}_N^J/\bar{a}_N^J$ will be bounded. It is  possible to
reverse our line of argument taking us from a 
Stieltjes function to the recurrence relation~\refeq{e_paderecrel} for
its Pad\'e numerators and denominators. According to Favard's theorem,
a series of coefficients $\{\bar{a}_n,\bar{b}_n\}$ with bounded
$1/\bar{a}_N^J$ and $\bar{b}_N^J/\bar{a}_N^J$ determines a unique 
compactly supported measure $d\phi(u)$, and a system of orthogonal
polynomials with respect to this measure, such that these coefficients
appear in their recurrence relation. The measure, in turn, determines
the Stieltjes function. 


All the two-point asymptotic functions $f(s)$ which appear in the
conformal approximation (which in QCD applies to the leading and
subleading perturbative contributions) are of the form
$f_n(s)=-(s-\lambda)^n 
\log(\lambda-s)$, with $n$ a positive integer or zero, or
$f_\nu(s)=(-1)^{[\nu]+1}(\lambda-s)^\nu$ with $\nu$ a positive
non-integer real number and $[\nu]$ the
integer closest to $\nu$ with $[\nu]\leq \nu$ (the entire part for
positive $\nu$). The variable $s$ is related to $t=p^2/\mu^2$
by $s=t+\lambda$. These functions are analytic at $s=0$, and
their Maclaurin 
series have radius of convergence $\lambda$. On the other hand, the
divergent behaviour at large $|s|$ can be taken care of by performing
$n+1$ or $[\nu]+1$ subtractions, respectively. The subtracted
functions are Stieltjes functions, and hence
their Pad\'e approximants satisfy all the 
properties we have just derived. For instance,
for $f_1(s)=(1-s)\log(1-s)$, we need two subtractions: 
\bea
\bar{f}_1(s) &=& f_1(s)-f_1(0)-s f_1^\prime(s) \nn
&=& (1-s) \log(s-1)+s \nn
&=& s^2 \int_0^1 \frac{1-u}{1-su} \, du .
\eea
So, $\bar{f}_1/s^2$ is a Stieltjes function, analytic inside the open
circle of radius 1 centered at $s=0$. This equation is nothing but
\refeq{e_substraction} with $\bar{f}_1(s)=F^J(s^{-1})$ and
$J=2$. Therefore, the function 
\beq
\Pi_N(s) = f_1(0) + s f^\prime_1(0) + s^2 [N-1/N]_{\bar{f}_1}(s)  ,
\eeq
is exactly the same as the Pad\'e approximant $[N+1/N]_{f_1}(s)$. It
is clear that this is generalized to $J=m-1$ in the case of $m$
subtractions. 

Consider any $f_\nu(s)$ with real $\nu$ (possibly integer) and compute
the $[N+J/N]$ as above, with $J\geq [\nu]-1$. Changing variables to
$x=2w-1=(1-t)/(1+t)$ we find a weight $W^J(x)=(1-x)^\nu
(1-x)^{(J-\nu)}$ with $x\in (-1,1)$. Therefore,
$\pi^J_N(x)=P_N^{(\nu,J-\nu)}$ and
$\rho^J_N(x)=\bar{P}_N^{(\nu,J-\nu)}$, and the Pad\'e approximant is
\beq
[N+J/N]_{f_\nu} = \sum_{j=0}^J (f_\nu)_j s^j + \frac{s^J
  \bar{P}_N^{(\nu,J-\nu)} \left(\frac{2-s}{s}\right)}
 {P_N^{(\nu,J-\nu)} \left(\frac{2-s}{s}\right)} \, .
\eeq
From their integral representation given by~\refeq{e_numdef}, the
associated Jacobi polynomials can be written in terms of
hypergeometric functions. Note that we should only consider
functions $f_\nu$ with $\nu>-1$. In fact, when $\nu\leq -1$, the
function $f_\nu$ is too divergent at $s=\lambda$ to be a Stieltjes
function. Of course, the Pad\'e approximant can still be calculated,
but it will not share the good (physical) properties we have derived
in this section.

\section{Migdal's limit}
\label{s.ml}

Let us study the Migdal limit of the Pad\'e recurrence
relation~\refeq{e_paderecrel}. 
We take $\lambda=1$ and call $\tau=\sqrt{t}$
(with $\tau$ positive for positive $t$). 
the Migdal limit is $N\rightarrow \infty$ and $\tau\rightarrow 0$ with
fixed $J$ and $\tilde{\tau} = \tau N$. To find the limit
of equation~\refeq{e_paderecrel}, we write (suppressing the index $J$)
$T_N(s)=T(N,\tau)$, treat $N$ as a continuous variable and expand 
\beq
T_{N\pm1}(N,\tau) = T(N,\tau) \pm \frac{\partial}{\partial N} T(N,\tau) + \frac{1}{2}
\frac{\partial^2}{\partial N^2} T(N,\tau) + \ldots. 
\eeq
Then, keeping only terms up to two derivatives, we get a differential
equation of the form 
\beq
\label{e_intermediate}
\left[\frac{\partial^2}{\partial N^2} + 2 \frac{1-c_N (1+\tau^2)^2}{1+c_N
    (1+\tau^2)^2} \frac{\partial}{\partial N} + 2
    \frac{1-a_N-b_N (1+\tau^2)+c_N(1+\tau^2)^2}{1+c_N
    (1+\tau^2)^2}\right] T(N,\tau) = 0. 
\eeq
Let us assume now that the coefficients of the recurrence relation can
    be expanded at large $N$ as 
\beq
a_N = a^{(0)} + a^{(1)} \frac{1}{N} + a^{(2)} \frac{1}{N^2} + \ldots,
\eeq
and similarly for $b_N$ and $c_N$. Then, the second order differential
equation~\refeq{e_intermediate} has a finite non-trivial
Migdal limit if and only if the following conditions are
    met: 
\begin{align}
& c^{(0)} = 1 , \nn
& 2 - a^{(0)} - b^{(0)} = 0 , \label{e_migdalconditions} \\
& c^{(1)}-a^{(1)}-b^{(1)} = 0 . \nonumber
\end{align}
As long the norm of the orthogonal polynomials can also be expanded at
large $N$ (with a finite number of terms with positive powers of $N$),
the first condition is satisfied. The second condition is ensured by
Rakhmanov's theorem~\cite{Rakhmanov}: if the measure $d\phi$ is
supported in $[-1,1]$ and $\phi^\prime>0$ almost everywhere in
$[-1,1]$, then it belongs to the Nevai class\footnote{Measures in the
Nevai class are those with finite limits $\bar{a}_n\rightarrow
\bar{a}^{(0)}$, $\bar{b}_n\rightarrow \bar{b}^{(0)}$~\cite{Nevai}.}
with $\bar{a}^{(0)}=2$ and $\bar{b}^{(0)}=0$. The assumptions of 
the theorem are fulfilled by the measure of Stieltjes functions, in
the variable $x=2\lambda^{-1} w 
-1$. Even though the coefficients of the recurrence relation in this
variable are different, when $\lambda=1$ (for
which~\refeq{e_migdalconditions} apply), the value of $a_n+b_n$ is
unchanged.  On the other hand,
$a^{(0)}=\bar{a}^{(0)}$, $b^{(0)}=\bar{b}^{(0)}$ for the class of
norms just mentioned. The third condition is more restrictive. Even if
it is norm dependent, it cannot be adjusted without spoiling our
assumption that the coefficients can be expanded in $1/N$. We check
explicitly below that this condition is fulfilled by Jacobi polynomials.

If all three conditions are met, we find for $N\rightarrow  \infty$
the following second-order differential equation:  
\beq
\label{e_migdaldiffeq}
\left[\frac{d^2}{d\tilde{\tau}^2} - c^{(1)} \frac{1}{\tilde{\tau}}
  \frac{d}{d\tilde{\tau}} + 
  (2-b^{(0)}) + (c^{(2)}-a^{(2)}-b^{(2)}) \frac{1}{\tilde{\tau}^2} \right]
T(\tilde{\tau}) = 0. 
\eeq

In the case of standardized Jacobi polynomials $P^{(\alpha,\beta)}_n$
with any $\alpha$ and 
$\beta$ we have, going back to the variable $w$, $a^{(0)}=4$,
$a^{(1)}=-2$, $b^{(0)}=-2$, $b^{(1)}=1$, $c^{(0)}=1$ and $c^{(1)}=-1$
so we see explicitly that all the
conditions~\refeq{e_migdalconditions} are directly
satisfied. In this case, \refeq{e_migdaldiffeq} reads 
\beq
\label{e_besseldiffeq}
\left[\frac{d^2}{d\tilde{\tau}^2} + \frac{1}{\tilde{\tau}} \frac{d}{d\tilde{\tau}} + 4 -
  \frac{\alpha^2}{\tilde{\tau}^2} \right]  T(\tilde{\tau}) = 0. 
\eeq
This is a Bessel equation, with general solution
\beq
\label{e_besselsol}
T(\tilde{\tau})=C_1 J_\alpha(2 \tilde{\tau}) + C_2 Y_\alpha(2 \tilde{\tau}) .
\eeq
Actually, in the more general case of Eq.~\refeq{e_migdaldiffeq} we
  can write $T(\tilde{\tau})=\tilde{\tau}^{(1+c^{(1)})/2}
  V(\tilde{\tau})$ and rescale the variable to
  $\tilde{\sigma}=\left(\sqrt{2-b^{(0)}}/2\right) \tilde{\tau}$. Then,
  $V(\tilde{\sigma})$ obeys~\refeq{e_besseldiffeq} with $\tilde{\tau}
  \rightarrow \tilde{\sigma}$ and
  $\alpha^2=a^{(2)}+b^{(2)}-c^{(2)}+\left(1+c^{(1)}\right)^2/4$. 
Note that the common factor $\tilde{\tau}^{(1+c^{(1)})/2}$ will cancel
out in the quotient $R/S$. Actually, this factor comes from the
normalization of the orthogonal polynomials. On the other hand, the
rescaling of 
  $\tilde{\tau}$ amounts to a rescaling of $\tilde{\mu}$. Therefore,
  we see that, as far as the Migdal limit is concerned, and if the
  limit exists, it is
  sufficient to consider the limit of Jacobi polynomials and work with
  equation~\refeq{e_besseldiffeq}.

When the third condition is not fulfilled, the recurrence relation
does not have a good Migdal limit. In these cases, one could still try
to find a continuous differential equation by modifying the way in
which the limit is taken, and this was actually done (in a different
language) in Migdal's original paper~\cite{Migdal}. In the following
we consider only the simplest case in which the
Migdal limit, as defined here, is finite. We have seen that this
reduces to studying the Pad\'e approximants of
``conformal'' functions $f_n$ and $f_\nu$.

\section{The AdS/Migdal correspondence}
\label{s.mc}

In this section, we describe the general relation between the Migdal
approximation and extra dimensions, for two-point functions. We give a
simple argument showing that the Migdal limit unavoidably gives a
result which corresponds to a holographic calculation in certain 5D
geometries. The converse result does not always hold: in some cases
the holographic results cannot be obtained from a Migdal approximation
to their 
asymptotic Euclidean functions. In order to simplify the notation we
consider correlators of scalar operators, and comment at the end on
the generalization to higher spins.

We start with the field theory (Migdal) side, and show that
analyticity\footnote{The numerators and denominators are analytic in
$s$ since they are finite limits of polynomials, and hence convergent 
Taylor series at $s=0$.} and the condition of finite the Migdal limit,
together with information about the deep Euclidean 
limit and the leading infrared behaviour, completely fix the
two-point function $\Pi_\mathrm{Migdal}(t)$. We have seen at the end
of the previous 
section that a good  Migdal limit necessarily gives numerators $R$ and
denominators $S$ which satisfy the differential
equation~\refeq{e_besseldiffeq}. Therefore, $R$ and $S$ have the
form~\refeq{e_besselsol}, with coefficients $C_{1,2}$ which are
functions of $\tau$.\footnote{We are treating $\tau=p/\mu$ and
  $\tilde{\tau}=p/\tilde{\mu}$ as 
independent variables. In the following we also need to use the
behaviour in the variable $p^2$, for fixed $\mu,\tilde{\mu}$.}

Now, let us impose the asymptotic value of the two-point function. The
limit $|\tilde{\tau}|\rightarrow \infty$ is 
equivalent to $N\rightarrow \infty$ with fixed $\tau$. Hence, for any
non-positive $\tau^2$, the Pad\'e
approximant converges to $f(\tau^2+1)$. Therefore, the Migdal
limit of the Pad\'e approximant must have the form
\beq
\frac{R}{S} = \frac{f(\tau^2+1) J_\mu(2
  \tilde{\tau}) + A(\tau) 
  H_\mu^{(1)}(2 \tilde{\tau})} {J_\mu(2 \tilde{\tau}) +
  B(\tau) H_\mu^{(1)}(2 \tilde{\tau})} ,
\eeq
with $H_\mu^{(1)}$ the first Hankel
function, which goes to zero exponentially for large positive
imaginary part of the argument. 

Next, we impose that $R$ and $S$ be analytic in $p^2$ at $p^2=0$. This
fixes the functions $A(\tau)$ and
$B(\tau)$. Recall that for integer $n$, $J_n(z)$ is an entire function
and $Y_n(z)$ equals $2/\pi\, \log z$ plus an entire function, whereas
for any $\mu$, $z^{-\mu} J_\mu(z)$ is entire. Consider first
$f_n(s)=-\tau^{2n} \log(-\tau^2)$. Then, the index of the Bessel
functions must be an integer, and
\beq
\frac{R}{S}= \frac{-\tau^{n-m} \log(\tau^2) J_{n+m}(2
  \tilde{\tau}) + \pi \tau^{n-m} Y_{n+m}(2 \tilde{\tau})}
      {\tau^{-n-m}J_{n+m}(2 \tilde{\tau})} .
\eeq
On the other hand, for $f_\nu(s)=(-1)^{[\nu]+1}(-\tau^2)^\nu$ and
  $\nu$ a non-integer 
real, analyticity requires $\mu=\nu+m$ with integer $m$, and
 \beq
\frac{R}{S}= \frac{(-1)^{[\nu]+1} \tau^{\nu-m} J_{-\nu-m}(2
  \tilde{\tau})} {\tau^{-\nu-m} J_{\nu+m}(2 \tilde{\tau})}. 
\eeq

Finally, we use the fact that the Pad\'e approximants have no poles or
zeros at $p^2=0$, and assume that this still holds in the Migdal
limit.\footnote{This may be proven using the asymptotic
  distribution of zeros of the orthogonal polynomials.} The leading 
  behaviour at small $p^2$  
is $R/S\sim p^{-2m}$ in all cases. Hence, we see that $m=0$ and this
  gives the final result for $R/S$. Now, let us define the Migdal
  two-point function as 
  $\Pi_N^{J,l}=t^{-l} [N+J/N]_{t^l f}$, with integer
  $l$. As discussed 
  above, we should only consider $l>-1-\nu$. Since $R/S$
  has no poles or zeros at $p^2=0$, $\Pi^l_{\mathrm{Migdal}}$ will
  have a zero of 
  degree $-l$ if $l<0$, and a pole of degree $l$ if $l>0$. On the other
  hand, the asymptotic behaviour is independent of $l$. So, we can use
  the result above with $m=l$ and find 
\beq
\Pi_\mathrm{Migdal}^l = \frac{-t^n \log(t) J_{n+l}(2
  \sqrt{\tilde{t}}) + \pi t^n Y_{n+l}(2 \sqrt{\tilde{t}})}
      {J_{n+l}(2 \sqrt{\tilde{t}})} ,
\eeq
 \beq
\Pi_\mathrm{Migdal}^l= \frac{(-1)^{[\nu]+1} t^{\nu} J_{-\nu-l}(2
  \sqrt{\tilde{t}})} { J_{\nu+l}(2 \sqrt{\tilde{t}})}, 
\eeq
in the integer and non-integer cases, respectively. Conversely, we see
that we can reproduce a low-energy behaviour $\sim p^{-2l}$ in the
two-point function simply by choosing $l$ in the definition of the Migdal
two-point function.

We turn now to the holographic calculations in 4+1 dimensions. 
In order to keep 4D Poincar\'e invariance, the geometry must be a warped
direct product of Minkowski times a one dimensional space $I$, which can
be chosen flat. In order to have a discrete spectrum, $I$ must be
compact. As long as the warp
factor is strictly monotonic, one can define coordinates in which the
metric is manifestly conformally flat:
\beq
ds^2 = \xi(z)^2 \left(dx^\mu dx_\mu - dz^2 \right) .
\eeq
In the coordinate $z$, $I=[z_{\mathrm{UV}},z_{\mathrm{IR}}]$. AdS geometry corresponds to
$\xi(z)=(kz)^{-1}$; in this case, the UV (IR) boundaries at $z_{\mathrm{UV}}$
($z_{\mathrm{IR}}$) hide the AdS boundary (horizon) at $z=0$ ($z=\infty$). 
The holographic prescription to calculate correlation functions of
field-theory operators at large $N$ (and strong t'Hooft coupling) is
given by the AdS/CFT correspondence~\cite{adscft}: calculate the
value of the action for on-shell bulk fields with fixed UV
boundary values, which act as sources for the dual
operators; then, differentiate functionally with respect to the
sources and put them to zero. For two-point functions, the on-shell
action reduces to a boundary term. For scalars,
\beq
\label{e_holographicformula}
\Pi(p) = \lim_{z_{\mathrm{UV}}\rightarrow 0} \left\{ X \left[\frac{\partial_z
    G(z,p)}{G(z,p)}\right]_{z\rightarrow {z_{\mathrm{UV}}}} +
\mathrm{counterterms} \right\} ,
\eeq
where $G(z,p)$ is the bulk-to-boundary propagator in a mixed
    position-momentum representation, fulfilling some specified
boundary conditions on the IR  boundary and free on the UV.
$X$ is a $p$-independent factor which cancels a divergent
factor in the non-analytic part in the limit $z_{\mathrm{UV}}\rightarrow 0$. The
    remaining 
counterterms in~\refeq{e_holographicformula} form a polynomial in
$p^2$ which cancels the poles at $z_{\mathrm{UV}}=0$. The propagator $G$
    satisfies the equation of 
    motion for the dual field. We can write a generic IR
    boundary condition as 
    $\hat{G}(z_{\mathrm{IR}},p)=0$, with $\hat{f}(z,p)= \kappa_1(p^2 z^2) f(z,p)+
    \kappa_2(p^2 z^2) z \partial_z f(z,p)$. Such a boundary condition
    can be obtained including mass, kinetic and higher-derivative
    terms localized on the IR boundary. Indeed, higher derivatives of
    $G$ in the variable $z$ can be
    written in terms of $G$ and $\partial_z G$ using the bulk equation
    of motion. Let $J$ and $Y$ be two independent solutions of the
    equation of motion. Then, we can write 
\beq
G(z_{\mathrm{IR}},p) = \hat{J}(z_{\mathrm{IR}},p) Y(z,p) - \hat{Y}(z_{\mathrm{IR}},p) J(z,p) \, .
\eeq
Taking the limit, the holographic formula in which
    the IR conditions are manifest has the form
\beq
\label{e_IRholographicformula}
\Pi(p) = \frac{A_1(p) \hat{J}(z_{\mathrm{IR}},p) - A_2(p) \hat{Y}(z_{\mathrm{IR}},p) }
    {A_3(p) \hat{J}(z_{\mathrm{IR}},p) - A_4(p) \hat{Y}(z_{\mathrm{IR}},p)} +
    \mathrm{local~terms} . 
\eeq
Our aim is to relate the differential equation satisfied by the limit
    of the Pad\'e numerator and denominator to a differential equation
    for $\hat{K}(z_{\mathrm{IR}},p)$ in the variable $z_{\mathrm{IR}}$, where $K$ is any
    linear combination of $J$ and $Y$. Then,
    meromorphicity, which follows from the discreteness of the
    spectrum in a compact space, will imply that the holographic
    two-point function will be the same as the Migdal one for some
    value of $l$. The function $f$ in the Migdal approach
    corresponds to the limit $z_{\mathrm{IR}}\rightarrow \infty$
    of~\refeq{e_holographicformula}, with $p^2<0$. In fact, for the
    functions we are 
    considering, which only depend on $p^2/\mu^2$, we should take as
    well a low-energy limit in which all the scales but $|p|$ (and
    $1/z_{\mathrm{IR}}$, which we have sent to zero) go to infinity. Then, the
    equivalence of the complete function will hold only in this limit
    (with finite $z_{\mathrm{IR}}$). 

In terms of the length variable $L=2/\tilde{\mu}$,
  Eq.~\refeq{e_besseldiffeq} reads 
\beq
\label{e_IRdiffeq}
\left[\frac{d^2}{dL^2} +  \frac{1}{L} \frac{d}{dL} +
  p^2 - \frac{\alpha^2}{L^2} \right] T(Lp/2) = 0 .
\eeq
Note that dimensional analysis plus the fact that $p^2$ appears only as
$p^2 T$ completely fix the form of this equation.
On the other hand, the equation of motion of a scalar of mass $M$ is
\beq
\label{e_eom}
\left[\partial_z^2+3 \left(\partial_z \log \xi(z) \right) \partial_z +
  p^2 - \xi^2(z) M^2 \right] \phi(z,p)=0 .
\eeq
In the case of Dirichlet boundary conditions, $\kappa_1=1$,
  $\kappa_2=0$, and 
$\hat{K}(z_{\mathrm{IR}},p)=K(z_{\mathrm{IR}},p)$ obeys the same equation as $\phi(z,p)$:
\beq
\label{e_IReom}
\left[\partial_{z_{\mathrm{IR}}}^2+3 \left(\partial_{z_{\mathrm{IR}}} \log \xi(z_{\mathrm{IR}}) \right)
  \partial_{z_{\mathrm{IR}}} + p^2 - \xi(z_{\mathrm{IR}})^2 M^2 \right] \hat{K}(z_{\mathrm{IR}},p)=0. 
\eeq
We see that, because we are using conformal coordinates, the
  normalization of the term with $p^2$ 
  in~\refeq{e_IReom} is the same as in~\refeq{e_IRdiffeq}. If $M\neq
  0$, complete agreement with the Migdal equation~\refeq{e_IRdiffeq}
  requires that $\xi(z)=(kz)^{-1}$, with $k$ a constant with
  dimensions of mass. Therefore, the space must be a slice of AdS with
  curvature $k$. In this case, $3 \partial_{z_{\mathrm{IR}}} \log \xi=-3/z_{\mathrm{IR}}$. To
  go to the normalization of~\refeq{e_IRdiffeq} we only need to write
  $\hat{K}(z_{\mathrm{IR}},p)=z_{\mathrm{IR}}^2 \hat{H}(z_{\mathrm{IR}},p)$. Then, for $z_{\mathrm{IR}}=L$ and
  $m^2\equiv M^2/k^2=\alpha^2-4$, the equation for $\hat{H}$ is
  exactly the same as 
  Eq.~\refeq{e_IRdiffeq}. From the AdS/CFT relation between the
  conformal dimension of the operator $\Delta$ and the mass of the
  dual field, we see that $\Delta=\alpha+2$, as it should (remember
  that $\alpha$ is the exponent of the asymptotic two-point function,
  which is determined by conformal invariance).
On the other hand, in the case $M=0$, we find agreement with Migdal
  equation for any $\xi(z)=(kz)^\eta$. Then, we define
  $\hat{K}(z_{\mathrm{IR}},p)=z_{\mathrm{IR}}^{(1-3\eta)/2} \hat{H}(z_{\mathrm{IR}},p)$ and
  identify $\alpha^2=(3\eta-1)^2/4$. Note that when $\alpha\neq 0$
  there are two different values of $\eta$ which give the same
  $\alpha$. Of course, for $\eta\neq -1$ the  
  geometry is not asymptotically AdS, and the
  AdS/CFT dictionary linking masses to conformal dimensions must be 
  modified. In the particular case $\eta=0$ we have flat space, and we
  see that the same Dirichlet two-point function, behaving
  asymptotically like $-(-t)^{1/2}$, is found in flat space with a
  massless scalar and in AdS with $m^2=-15/4$. At any
  rate, adjusting the mass parameter $m$ we can always
  reproduce Migdal's differential equation in AdS with Dirichlet
  boundary conditions in the IR. Moreover,
  Dirichlet boundary conditions give a correlator which (for scalars)
  has not a zero nor a pole at $p^2=0$. Indeed, we can rescale the
  field with a $z$-dependent factor such that the mass term in the
  equation of motion cancels. This shows that any zero-mode must be
  flat. But then, the Dirichlet condition forces it to vanish. This is
  true for any UV boundary condition. For UV Dirichlet (Neumann)
  conditions, this is telling us that the (inverse) two-point function
  does not have a pole at zero momentum.\footnote{For the relation
  between holographic correlators and connected correlation functions
  of the AdS theory, including the propagator, see the third reference
  in~\cite{adscftrs}.}
  Therefore, we conclude that IR Dirichlet boundary conditions
  correspond to $l=0$ in the Migdal approach. 

Consider now Neumann boundary conditions, $\hat{K}=\partial_z K$. As
long as $M=0$, and for any $\eta$,  $\hat{K}$ satisfies the following
second-order 
differential equation:
\beq
\label{e_Neumanneom}
\left[\partial_{z_{\mathrm{IR}}}^2+ \frac{3\eta}{z_{\mathrm{IR}}} 
  \partial_{z_{\mathrm{IR}}} + p^2 - \frac{3\eta}{z_{\mathrm{IR}}^2} \right]
  \hat{K}(z_{\mathrm{IR}},p)=0.  
\eeq
Writing $\hat{K}=z_{\mathrm{IR}}^{(1-3\eta)/2} \hat{H}$, we reproduce Migdal
equation~\refeq{e_IRdiffeq} for $\alpha^2=(3\eta+1)^2/4$. Thus,
$\alpha_\mathrm{Dirichlet}^2-\alpha_\mathrm{Neumann}^2=-3\eta$. For
  positive $\alpha$ and $\eta\leq-1/3$,
  $\alpha_\mathrm{Dirichlet}-\alpha_\mathrm{Neumann}=1$.
Furthermore, in this case there would be a zero mode if Neumann boundary
conditions were also used in the UV, so the two-point
function has a simple zero. Therefore, IR Neumann boundary conditions
in a massless theory corresponds to choosing $l=-1$ in the Migdal
  approach. On the other hand, if $M \neq 0$, it is possible to write
  a second-order differential equation for $\hat{K}$, but with
  coefficients which are not analytic in $p^2$. So, this equation is
  not of the Migdal form and we cannot reproduce the holographic
  function within the Migdal approach.

Let us study next mixed boundary conditions with $\kappa_1=1$ and
$\kappa_2$ a constant, and assume $\eta=-1$ to start 
with. Then, the function $\hat{K}$ satisfies a differential equation of the  
Migdal form if and only if
$\kappa_2=\frac{2\pm\sqrt{4+m^2}}{m^2}$. After
rescaling, we find $\alpha^2=5+m^2 \mp 2\sqrt{4+m^2}$, respectively. 
It turns out that this boundary condition, which can be understood as
arising from a mass term localized on the IR brane, is automatic when the
scalar field is the supersymmetric partner of a fermion or a gauge
boson~\cite{GP}. In this case, a fine-tuned mixed UV boundary
condition arises as well, such that a zero-mode results. Therefore, this
corresponds again, up to local terms, to a Migdal function with
$l=-1$. On the other hand, if $M=0$ and $\eta$ arbitrary,
$\kappa_2=1/(3\eta-1)$ and $\alpha^2=[3(\eta-1)/2]^2$. Note that the
remaining solution which would correspond to infinite $\kappa_2$, is the one
studied in the Neumann case.
  
It is also possible to reproduce any integer value of $l$ with
$l+\alpha_{\mathrm{Dirichlet}}>-1$ by choosing adequate analytic
functions 
$\kappa_1$ and $\kappa_2$. This can be proven showing that a good
differential equation for $\hat{K}$ is obtained for discrete
values of the coefficients (depending on the mass) in the expansions
of $\kappa_1$ and 
$\kappa_2$, and studying their behaviour at small momentum. Describing
this in detail would be lengthy, so we simply observe that these properties
follow quite straightforwardly from the differential-recursion relations
of Bessel functions and leave the details to the interested reader. 

So, to summarize, every Migdal approximation of a two-point function
in the (finite) Migdal limit can be reproduced by an AdS calculation
with a given mass and fine-tuned boundary conditions. The Euclidean
asymptotic behaviour is determined by the mass, whereas the different
discrete choices of the parameter $l$, which controls the leading
behaviour at $p^2=0$, correspond to different discrete IR boundary
conditions in AdS. The converse is not true: not for any mass and IR
boundary condition can one find an equivalent Migdal approximation.
One can alternatively reproduce the Migdal calculations using a
different warped geometry when the mass of the dual field vanishes. In
all cases, the 
infrared Migdal parameter $\tilde{\mu}$ is proportional to the inverse
of the position of the IR boundary in the conformal coordinates
(the ones for which the metric is manifestly conformally flat).

All this discussion can be readily extended to higher
integer spins. Ultraviolet conformal invariance determines the form of
the two-point function in the UV as 
\beq
\la O_i O_j \ra = Z_{ij}(p)/p^{2n} f(p^2+1) ,
\eeq
where $i,j$ represent Lorentz indices and $Z_{ij}$ is a tensor,
polynomical in $p_\mu$. We choose $n$ such that $Z_{ij}/p^{2n}$ be
adimensional. For instance, for a vector operator of conformal
dimension $\Delta$, 
\beq
Z_{\mu\nu}=\left(\frac{2(\Delta-2)}{\Delta-1}
p_\mu p_\nu - \eta_{\mu\nu}p^2\right) 
\eeq
and $n=1$.
For a conserved current, $\Delta=3$ and we get a transverse function. 
We can directly apply Migdal's approximation to
$f$, and define
\beq
\Pi_{\mathrm{Migdal}\,ij}^l =
Z_{ij}(p)/p^{2n}\Pi_{\mathrm{Migdal}}(p^2)^l . 
\eeq
This keeps the tensorial form dictated by conformal
invariance. On the other hand, the holographic calculation will
preserve the conformal tensor structure
if the geometry is that of a slice of AdS. It will also preserve this
form for any metric in the case of completely antisymmetric tensors
which are dual to $p$-forms, due to gauge
invariance. Therefore, the results we have obtained for 
massless scalars can be generalized to higher $p$-forms, and in
particular to gauge fields. For instance, the unphysical example in
Section~\ref{s.mae} with asymptotic behaviour $\Pi(p^2)\sim p$
corresponds to a gauge field in flat space. An AdS calculation with
a vector field with $m^2=-3/4$ (corresponding to conformal
dimension 5/2) and adequate boundary terms would
reproduce the scalar function $\Pi(p^2)$, but not a transverse
tensor. Finally, in extending the discussion for scalars to tensors,
one should also take into account that the coefficients in the second
term of the equation of motion~\refeq{e_eom} will be different,
which leads to a different normalization and to a different relation
between $m$ and $\alpha$ (or, equivalently, between $m$ and
$\Delta$). For vector bosons, the coefficient 3 is changed to 1 and
$\alpha_{\mathrm{Dirichlet}}=\sqrt{1+m^2}$.

\section{Deconstruction and holography}
\label{s.dh}

\begin{figure}[b]
\centering{
\begin{picture}(250,50)(0,-20)
\put(0,0){\fcolorbox{white}{white}{
    \SetWidth{0.5}
    \GOval(-5,0)(14,14)(0){0.882}}}
\put(37,0){\vector(-1,0){25}}
\put(50,0){\circle{25}}
\put(87,0){\vector(-1,0){25}}
\put(100,0){\circle{25}}
\put(137,0){\vector(-1,0){25}}
\put(150,0){$\dots$}
\put(200,0){\circle{25}}
\put(237,0){\vector(-1,0){25}}
\put(250,0){\circle{25}}
\put(287,0){\vector(-1,0){25}}
\put(-2,-25){$A_\mu^0$}
\put(48,-25){$A_\mu^1$}
\put(98,-25){$A_\mu^2$}
\put(198,-25){$A_\mu^{N-1}$}
\put(248,-25){$A_\mu^N$}
\put(20,-10){$U_1$}
\put(70,-10){$U_2$}
\put(220,-10){$U_N$}
\put(270,-10){$U$}
\end{picture} 
}
\label{f.moose}
\caption{The moose diagram for our deconstruction setup}
\end{figure}
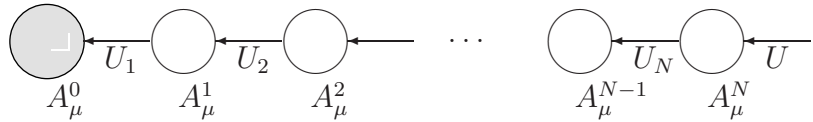

In the previous section we have studied the Migdal limit of Pad\'e
approximants. In this one, we relate the approximants with finite
$N$ to deconstruction models. In particular, this makes explicit the
mechanism by which Migdal correlators approach the holographic ones,
and how a discrete version of the holographic
formula~\refeq{e_holographicformula} is realized by the Pad\'e
approximants. For definiteness, we will stick to the case of gauge
bosons---dual to conserved currents---and discuss Dirichlet and Neumann
conditions only.

We consider a deconstruction model corresponding to the moose diagram
sketched in \fref{moose}.  
It involves a chain of $N$ $SU(N_F)$ groups with the gauge fields 
$A_\mu^j = A_\mu^{j,a} T^a$,  $j= 1 \dots N$.
The groups communicate with nearest neighbours via bifundamental non-linear sigma model field $U_j$, referred to as the  links.  
They are unitary $N_F \times N_F$ matrices with determinant equal
to $1$.  
The role of the last non-linear sigma field $U$ is to control the boundary conditions for the gauge field (the analogue of IR boundary conditions in 5D).

At the left end of the chain we singled out the ``boundary'' gauge field  $A_\mu^0$.
Similarly as in AdS/CFT, this boundary field is interpreted as an external current probing the dynamics of the ``bulk'' model. 
Here, the bulk refers to the remaining gauge fields $A_\mu^j$, $j\geq 1$.
The latter will show up as resonances in the boundary field correlation function. 

In the following  we calculate the two-point correlation function of  the boundary  fields in deconstruction.
We do it first in the more familiar minimal deconstruction set-up, and
then in what we call tilted deconstruction, which contains additional
interactions between the neighbouring sites.   
Correlators obtained in tilted deconstruction turn out to be directly related to Migdal's Pad\'e approximants and we work out a dictionary between the two approaches.   

\subsection{Minimal deconstruction}

The gauge transformations $\omega_j$, $j = 0 \dots N$, act  as 
$A_\mu^j \to\omega_j A_\mu \omega_j{}^\dagger - i \pa_\mu \omega_j \omega_j{}^\dagger$ 
and 
$U_j  \to \omega_{j-1} U_j \omega_j{}^\dagger$. 
The simplest non-trivial action that is invariant under these transformations can  be written as
\beq
\label{e.da} 
S =  \int d^4 x  \sum_{j=1}^{N}  \left ( 
-{1 \over 2 g_j^2}  {\rm tr} \{F_{\mu\nu}^j   F_{\mu\nu}^j  \} 
+ {\rm tr}  \{ v_j^2 D_\mu U_j D_\mu U_j^\dagger  \} 
\right ) +  \int d^4 x {\rm tr}  \{ v^2 D_\mu U D_\mu U^\dagger  \} \, ,
\eeq
with $D_\mu U_j  = \pa_\mu U_j - i A_\mu^{j-1} U_j + i U_j A_\mu^{j}$, 
$D_\mu U  = \pa_\mu U - i A_\mu^{N} U$.
This action is minimal in the sense that the only interactions between various gauge fields come from the covariant derivatives acting on the links. 
The relation of this deconstruction setup to 5D gauge theories can be worked out analogously as in \cite{faki}. 
Consider the 5D action for a gauge field in the background  $ds^2 = a^2(z) dx^2 - b^2(z) dz^2$: 
\beq
S_{5D} =  \int d^4 x \int_0^L dz \sqrt{g} 
\left ( -{1 \over 2 g_5^2}  {\rm tr} F_{MN}^2   \right ) 
 \to 
\int d^4 x \int_0^L dz  
\left ( -{ b(z) \over 2 g_5^2} {\rm tr} F_{\mu\nu}^2   + {a^2(z) \over  g_5^2 b(z)}  {\rm tr}(\pa_5 A_\mu)^2  
\right ) \, .
\eeq
Latticizing the  5th coordinate,  $z \to z_j = j \Delta$, 
$\pa_5 f(z) \to (f(z_{j}) - f(z_{j-1}))/\Delta$ we obtain:   
\beq
S_{5D}  \to 
\int d^4 x \sum_{j = 1}^N 
\left ( -{b(z_j) \Delta \over 2 g_5^2} {\rm tr} F_{\mu\nu}^2(z_j)   
+ {a^2(z_j) \over  g_5^2 \Delta b(z_j)}  {\rm tr}
(A_\mu(z_j) - A_\mu(z_{j-1}))^2  
\right ) \, .
\eeq 
This can be mapped onto the deconstruction action  \erefn{da}  (in the unitary gauge $U_j = 1$). 
The warp factors and the lattice spacing  translate into the parameters of the deconstruction lagrangian according to the following dictionary:  
\beq
\label{e.dic}
a(z_j) \to {v_{j} \over v_1} {g_1 \over g_j} \, ,
\qquad
b(z_j) \to {g_1^2 \over g_j^2} \, ,
\qquad 
\Delta \to {1 \over g_1 v_1} \, ,
\qquad
g_5^2 \to  {g_1 \over v_1}   \, .
\eeq
We have fixed $a(z_1) = b(z_1) = 1$. 
In passing we note that discretization in Poincar\'e coordinates,
$b(z)=1$,  corresponds to $g_j = g$, while  
discretization in conformal  coordinates, $b(z)=a(z)$, corresponds to $g_j v_{j}= g v$. 

We can integrate out all the bulk gauge fields 
and obtain an effective action for the boundary field.
At tree-level, the integrating-out amounts to 1) solving the equations of motion for the bulk fields in the presence of a background boundary field and 2) inserting the solution into the deconstruction action.
The details of this procedure are given in the appendix.    
At the quadratic level, the effective action in the momentum space has the form
\beq
S_{eff} = \int {d^4p \over (2\pi)^4}   
v_1^2 A_\mu^0(p) \left (-\eta_{\mu\nu}+ {p_{\mu}p_{\nu} \over p^2}  \right ) A_\nu^0(p) \Pi (p^2) \, . 
\eeq  
The polarization function is given by a compact expression 
\beq
\label{e.pod} 
{F_N^1(p^2)  \over F_N^0(p^2)} 
+ {1 \over g_0^2 v_1^2} p^2 - 1 \, .
\eeq 
where $F_N^j$ is a solution to the equation of motion 
\beq
\label{e.dme}
\left ( v_{j+1}^2 + v_j^2 -  {p^2 \over g_j^2} \right ) F_{N}^j  
-  v_j^2 F_N^{j-1}  - v_{j+1}^2 F_N^{j+1} = 0 \, ,
\eeq
subject to a boundary condition at $j = N$.
The boundary condition is controlled by  the parameter $v$ in the lagrangian. 
The limit $v \to 0$ leads to a deconstructed analogue of the Neumann boundary condition, 
\beq
F_N^{N+1} = F_N^{N}  
\, ,
\eeq
while $v \to \infty$  corresponds to  the Dirichlet boundary condition, 
\beq
F_N^{N} = 0  
\, .
\eeq
Intermediate values of $v$ correspond to mixed boundary conditions. On
the other hand, for $g_0 \rightarrow \infty$ we recover the case of
non-dynamical boundary fields, which can be regarded as sources.

There are many apparent similarities between polarization functions obtained in minimal deconstruction  and those derived using the Migdal procedure.  
Let us point them out.  
\ben
\item 
The correlation function is represented as a ratio of two polynomials in $p^2$: 
\beq
\Pi(p^2) = {R_{N}(p^2)  \over  S_N(p^2)} \, ,
\eeq 
where 
\beq
R_N= F_N^1  + F_N^0 \left ( {p^2 \over g_0^2 v_1^2} - 1 \right )
\qquad
S_N =  F_N^0 
\eeq
Indeed, from the equation of motion \erefn{dme} $F_N^j$ is a polynomial of degree $N-j$ in $p^2$, once we set $F_N^{N} = {\rm const}$ (in the Dirichlet case $F_N^{N-1} = {\rm const}$ and $F_N^j$ has the degree $N-j-1$). 
\item 
The numerator and the denominator satisfy a second-order recurrence
relation in the degree $N$, and the equation is the 
same for both.
For example, in the Dirichlet case it is given by ($T = R,S$):  
\beq
\label{e.drr}
T_{N+1}(p^2)  = \left (1 + {v_N^2 \over v_{N+1}^2} -  {p^2 \over g_N^2 v_{N+1}^2} \right ) T_{N}(p^2)  
-  {v_N^2 \over v_{N+1}^2} T_{N-1}(p^2) \, . 
\eeq   
This follows from the fact that the solution $F_N^j$ satisfying the Dirichlet boundary conditions
can be written as  $F_N^j = Y_N J_j -  J_N Y_j$, where $J_j$ and $Y_j$ are any two independent solutions to \eref{dme}. 
\item In the limit $N \to \infty$ and for $|p^2| \ll v_1^2$ and
  $p^2<0$, the polarization function obtained in deconstruction
  approximates the non-analytic behaviour of the polarization function
  in the corresponding 5D model in the deep Euclidean regime.  
For example, in the deconstructed $\ads$ models one obtains \cite{blfape}
$\Pi(p^2) \sim  -p^2 \log (-p^2/v_1^2)$, while in the deconstructed flat
  models we find 
$\Pi(p^2) \sim  p^2 (-p^2/v_1^2)^{-1/2}$.  
\een

In spite of these similarities it is not possible to find a precise
mapping between the Migdal approximation and minimal deconstruction.  
The reason is that  the recurrence relations \erefn{drr} and (\ref{e_paderecrel}) are incompatible. 
Indeed, the form of the recurrence relation \erefn{drr} implies that $T_N(p^2)$ is an orthogonal polynomial in the variable $p^2$. 
On the other hand, the numerators and denominators obtained by the Migdal procedure, although related to orthogonal polynomials by \eref{mdopr}, are themselves {\em not\/} orthogonal polynomials.    

In the following we explore a modified deconstruction framework that allows for a mapping of the boundary correlators to those obtained using the Migdal approximation. lators to those obtained using the Migdal approximation. 

\subsection{Tilted deconstruction}

We modify the minimal deconstruction action \erefn{da} adding a kinetic mixing between neighbouring gauge fields,
\bea 
\label{e.tda} 
S =  & \int d^4 x  \sum_{j=1}^{N}  \left ( 
-{1 \over 2 g_j^2}  {\rm tr} \{F_{\mu\nu}^j   F_{\mu\nu}^j  \} 
+ {\rm tr}  \{ v_j^2 D_\mu U_j D_\mu U_j^\dagger  \} 
\right ) +  \int d^4 x {\rm tr}  \{ v^2 D_\mu U_{N+1} D_\mu
U_{N+1}^\dagger  \}  
\nl 
- \int d^4 x  \sum_{j=1}^{N} {\alpha_j \over  2g_j^2} {\rm tr}  \{ 
F_{\mu \nu}^{j-1} U_j  F_{\mu \nu}^j U_j^\dagger  + \hc \} 
+ \int d^4 x {1 \over 2 g^2}  {\rm tr} \{F_{\mu\nu}^N   F_{\mu\nu}^N \}  
\eea
Such deconstruction setup is also related to a latticized 5D gauge theory in the warped background. 
The difference with the minimal case is that the 5D action must contain a higher derivative term breaking the 5D Lorentz invariance: 
\beq
S_{5D} \to 
\int d^4 x \int_0^L dz  
\left ( -{ b(z) \over 2 g_5^2} {\rm tr} F_{\mu\nu}^2   
+ {a^2(z) \over  g_5^2 b(z)}  {\rm tr}(\pa_5 A_\mu)^2 
-  {\alpha(z) \over  2 g_5^2}   {\rm tr} F_{\mu\nu} \pa_z^2  F_{\mu\nu}
\right ) \, .
\eeq
The dictionary between deconstruction and 5D is now given by: 
\bea &
\label{e.tdic}
a(z_j) \to {v_{j+1} \over v_1} {g_1 \over g_j} {\sqrt{1 - \alpha_j} \over \sqrt{1 - \alpha_1}} \, , 
\qquad
b(z_j) \to {g_1^2 \over g_j^2} {1 - \alpha_j \over 1 - \alpha_1}  \, ,
\qquad 
\alpha(z) \to {\alpha_j \over g_j^2 v_1^2}  \, ,
\nl 
\Delta \to {\sqrt{1 - \alpha_1} \over g_1 v_1} \, ,
\qquad
g_5^2 \to  {g_1 \over v_1 \sqrt{1- \alpha_1} }   \, .
&\eea
We see that tilted deconstruction corresponds to a 5D setup with
some specific higher-derivative terms. Nevertheless, the extra term is
irrelevant at low energies and a standard
lowest-order 5D action is recovered. Hence, the effect of the mixing term
amounts to a renormalization of the coefficients in this action.
Viewed as  a 4D field-theoretical model, tilted deconstruction is
healthy as long as the mixing coefficients $\alpha_j$
are not too large (so that there are no ghosts). 

As in the minimal setup, it makes sense to integrate out the resonances $A_\mu^j$, $j \geq 1$ and calculate the effective action for $A_\mu^0$. 
The polarization function is given by (see \aref{ba} for a derivation)    
\beq
\label{e.tpod} 
\Pi(p^2) = 
{F_N^1(p^2)  \over F_N^0(p^2)} \left ( 1 + {\alpha_1 \over g_1^2 v_1^2} p^2 \right ) 
+ {1 \over g_0^2 v_1^2} p^2 - 1 \, .
\eeq 
The last two terms have a form of a local polynomial in $p^2$. 
By adding higher derivative terms for the boundary fields to the deconstruction action we could, in fact, obtain an arbitrary polynomial in $p^2$.  

In tilted deconstruction, $F_N^j$ solve a modified equation of motion 
\beq
\label{e.tdme}
\left ( v_{j+1}^2 + v_j^2 -  {p^2 \over g_j^2} \right ) F_{N}^j  
-  \left ( v_j^2 + {p^2 \alpha_j \over g_j^2} \right )F_N^{j-1}  
- \left (v_{j+1}^2 + {p^2 \alpha_{j+1} \over g_{j+1}^2} \right ) F_N^{j+1} = 0 \, ,
\eeq
subject to the boundary condition 
\beq
\left ( v_{N+1}^2 - v^2 + {1 \over g^2}p^2 \right ) F_N^{N}
= \left (v_{N+1}^2  + {\alpha_{N+1} \over g_{N+1}^2} p^2  \right
)F_N^{N+1} \, .
\eeq
In the limit $v \to \infty$ we obtain Dirichlet boundary conditions, 
$F_N^N = 0$,   
while setting $v = 0$, $\alpha_{N+1}/g_{N+1}^2 = 1/g^2$ leads to Neumann boundary conditions, 
$F_N^{N+1} = F_N^N$. 

We will prove that, for certain choices of the coefficients $\alpha_j$, the polarization functions obtained in this setup are directly related to those obtained by Migdal approximation. 

\subsection{Migdal--deconstruction map}

Let us make the following ansatz for the mixing coefficients: 
\beq
{\alpha_j \over g_j^2} = {v_j^2 \over \mu^2} \, ,
\eeq
where $\mu$ is an arbitrary scale.
We also introduce a new variable,  $s = 1 + p^2/\mu^2$. 
The equation of motion now becomes 
\beq
\label{e.tadme}
\left ( v_{j+1}^2 + v_j^2 +  {\mu^2 \over g_j^2}  - s {\mu^2 \over g_j^2} \right ) F_{N}^j(s)  
- v_j^2 s F_N^{j-1}(s)   - v_{j+1}^2 s F_N^{j+1}(s) = 0 \, .
\eeq
We find  it convenient to discuss the Dirichlet and the Neumann case
separately.  

\bc 
{\em Dirichlet boundary conditions} 
\ec

We investigate the solutions to \eref{tadme} subject to the boundary
condition $F_N^N=0$.  
If we set $F_N^{N-1} = {\rm const}$ then 
$s^{j} F_N^{N-j-1}(s)$ is a polynomial in $s$ of degree $j$. 
We define the polynomials 
\beq
\label{e.dndd}
R_N^1(s) = s^{N+1} F_{N+1}^1(s),
\qquad 
\qquad 
S_N^1(s) = s^{N} F_{N+1}^0(s).
\eeq   
It follows that $S_N$ has degree $N$, while $R_N$ has degree $N+1$. 
From \eref{tpod} the polarization function can be written as 
\beq
\Pi(s) = {R_N^1(s) \over S_N^1(s)}  +  f_0 +  f_1 s 
\eeq
and has the form of a Pad\'e approximant with $J = 1$. 
Indeed, the numerator and the denominator as defined in \eref{dndd}
satisfy the second order difference equation  
\beq
\label{e.tdrr}
T_{N+1}^1(s)  = \left (1 + {v_{N+1}^2 \over v_{N+2}^2} 
+ {\mu^2 \over  g_{N+1}^2 v_{N+2}^2}  - {\mu^2 \over  g_{N+1}^2 v_{N+2}^2} s  \right ) T_{N}^1(p^2)  
-  {v_{N+1}^2 \over v_{N+2}^2} s^2 T_{N-1}^1(s) \, . 
\eeq   
subject to the boundary conditions 
\beq
\label{e.tdrrbc}
R_{0}^1(s) = 0, \quad R_{1}^1(s) = \mathrm{constant}, \quad 
S_{-1}^1(s) = 0, \quad S_{0}^1(s) = \mathrm{constant} 
\eeq  
Equations \erefn{tdrr} and \erefn{tdrrbc} follow simply from the fact that 
the solution $F_{N+1}^j$ satisfying Dirichlet boundary conditions can be written as 
$F_{N+1}^j = Y_{N+1} J_j -  J_{N+1}Y_j$, where $J_j$ and $Y_j$ are any two independent solutions to \eref{tadme}.   
They have exactly the same form as the recurrence equations and the boundary conditions for the denominator and the numerator in Pad\'e approximation.

In fact, the recurrence relation in Pad\'e approximation contains three sets of coefficients 
$a_N$, $b_N$ and $c_N$, which define the related orthogonal polynomial.    
On the deconstruction side we dispose only of two sets: $g_N$ and $v_N$. 
However, since rescaling of the numerator and the denominator by the same function does not change the polarization function, Migdal approximation and deconstruction are equivalent if the recurrence equations can be brought to the same form after rescaling $T_N^1$ by an arbitrary function,  $T_N^1 \to h_N T_N^1$. 
Taking this into account leads to the norm-independent consistency
conditions: 
\bea
{g_{N+1}^2(v_{N+2}^2 + v_{N+1}^2) \over \mu^2 } &=& - 1  - {a_N \over b_N} 
\nn
{g_N^2 g_{N+1}^2 v_{N+1}^4 \over \mu^4 } & =& {c_N \over b_N b_{N-1}} 
\eea
Thus, given $a_N$, $b_N$ and $c_N$ defining the orthogonal polynomial
corresponding to the Pad\'e approximants, we are able to  reconstruct
the (tilted) deconstruction model that would give exactly the same
polarization function.  
Furthermore, using the dictionary \erefn{tdic} we can find the
continuum background.  

Let us now investigate what deconstruction model corresponds to the Pad\'e approximants found in section 2. 
Those examples where all associated with Jacobi polynomials $P_N^{\alpha,\beta}(2 s-1)$,
 whose recurrence relation involves the coefficients
\bea
a_N &=& { (2 N + \alpha + \beta + 1) (2 N + \alpha + \beta + 2) \over 
 (N+1)(N + \alpha + \beta + 1)} \, ,
\nn
b_N &=& { (2 N + \alpha + \beta + 1)(\alpha^2 - \beta^2 -  (2 N + \alpha + \beta )(2 N + \alpha + \beta + 2) )\over 
 2 (N+1)(N + \alpha + \beta + 1)(2 N + \alpha + \beta ) } \, ,
\nn
c_N &=& { (N + \alpha ) (N + \beta )(2 N + \alpha + \beta + 2) \over 
 (N+1)(N + \alpha + \beta + 1)(2 N + \alpha + \beta )} \, .
\eea
For large $N$ the consistency equations can be approximated by: 
\bea
{g_{N+1}^2(v_{N+2}^2 + v_{N+1}^2) \over \mu^2 } &=& 
1  +  {\alpha^2 - \beta^2 \over 2} {1 \over N^2} + \co(1/N^3)  \, , 
\nn
{g_N^2 g_{N+1}^2 v_{N+1}^4 \over \mu^4 }  &=& 
{1 \over 4} + {1 - 4 \beta^2 \over 16} {1 \over N^2} + \co(1/N^3).  
\eea 
They can be approximately solved by 
\bea
{v_{N+1}^2 \over v_{N}^2} &=& 1 + {1 \pm 2\alpha \over N} + \co(1/N^2) \, , 
\nn
{g_{N+1}^2 \over g_{N}^2} &=& 1 - {1 \pm 2\alpha \over N} + \co(1/N^2) \, ,
\nn
{g_N^2 v_N^2 \over \mu^2} &=& {1 \over 2} \left ( 1 - {1 \pm  2\alpha \over 2 N} + \co(1/N^2) \right ).
\eea
The corresponding deconstruction background, at lowest order, does not
depend on $\beta$.  
We see that $\alpha = 1/2$ can be matched to flat deconstruction
with $g_N$ and $v_N$ independent of $N$.  
On the other hand, $\alpha = 1$ is equivalent to deconstruction with 
\beq
\label{e.dadsb} 
v_N^2 \approx v_1^2 \frac{1}{N} \, , \qquad 
g_N^2 \approx {1 \over 2} g_1^2 N \, ,
\eeq 
which by \eref{tdic} is deconstruction of $\ads_5$  gauge theories
latticized in conformal coordinates.  

\bc 
{\em Neumann  boundary conditions} 
\ec

The results for the Neumann boundary conditions  $F_{N}^{N+1} = F_N^N$ can be obtained in an  analogous way, and  below we simply give our results. 
We define the polynomials:  
\beq
\label{e.dnd}
R_N^0(s) = {1 \over s - 1} s^{N}(s F_{N}^1(s) -  F_{N}^0(s))
\qquad 
\qquad 
S_N^0(s) = s^{N} F_{N}^0(s)
\eeq   
Both $R_N$ and $S_N$ are polynomials of degree $N$ in $s$ (one can show that the factor $1/(s-1)$ always cancels out). 
The polarization function can be written as 
\beq
\Pi(s) = {p^2 \over \mu^2} \left ( {R_N^0(s) \over S_N^0(s)}  + f_0  \right )
\eeq
and has the form of a Pad\'e approximant with $J = 0$. 
The polynomials satisfy a recurrence
relation that is different from the Dirichlet case:  
\beq
\label{e.tnrr}
T_{N+1}(s)  = \left ({v_{N+1}^2(g_N^2 + g_{N+1}^2) + \mu^2  \over g_{N+1}^2 v_{N+2}^2} 
- {\mu^2 \over  g_{N+1}^2 v_{N+2}^2} s  \right ) T_{N}(p^2)  
-  {g_N^2 v_{N}^2 \over g_{N+1} v_{N+2}^2} s^2 T_{N-1}(s) \, . 
\eeq
In obtaining this equation, the correlation between the coefficients
of~\refeq{e.tadme} is crucial. Adding a mass term would spoil this
correlation and the resulting recurrence relation in the Neumann case
would not be of the Pad\'e form. This agrees with our discussion of
the continuum extra dimensions.
The consistency conditions are given in this case by: 
\bea
{v_{N+1}^2(g_{N+1}^2 + g_{N}^2) \over \mu^2 } &=& - 1  - {a_N \over b_N}  
\nn
{g_N^4 v_N^2 v_{N+1}^2 \over \mu^4 } & =& {c_N \over b_N b_{N-1}} \, . 
\eea
For Jacobi polynomials on the Migdal side the large-$N$ approximate solution 
is given by 
\bea
{v_{N+1}^2 \over v_{N}^2} &=& 1 -  {1 \pm 2\alpha \over N} + \co(1/N^2) 
\nn
{g_{N+1}^2 \over g_{N}^2} &=& 1 + {1 \pm 2\alpha \over N} + \co(1/N^2) 
\nn
{g_N^2 v_N^2 \over \mu^2} &=& {1 \over 2} \left ( 1 + {1 \pm 2\alpha \over 2 N} + \co(1/N^2) \right ).
\eea 
We can see that $\alpha = -1/2$ is reproduced by flat deconstruction
with $g_N$ and $v_N$ independent of $N$, while $\alpha = 0$
corresponds again to deconstruction of $\ads_5$ in conformal
coordinates,  
\beq
\label{e.nadsb} 
v_N^2 \approx v_1^2 {1 \over N} \, , \qquad 
g_N^2 \approx {1 \over 2} g_1^2 N  \, .
\eeq 
These results are very welcomed, as they show that a single
deconstruction setting is able to reproduce the values $l=0$ and $l=1$
in the Migdal approximation, just by a change of the IR boundary
condition.

\section{Conclusions}
\label{s.c}

In this paper we have studied the relations between three different
methods of 
computing correlation functions in strongly-coupled large-$N_c$ 
theories: the Migdal 
approach via Pad\'e approximants, the 5D holographic approach via
boundary correlators, 
and the deconstruction approach via external field correlators.  We have
made explicit the connection between the Migdal approximation and the other
two methods. The key feature of the Pad\'e approximant is that its
denominator and numerator can be expressed in terms of some
orthogonal polynomial and their associated
orthogonal polynomial, respectively.
This ensures physical properties of the  Pad\'e 
approximants analogous to those of large-$N_c$ theories.
Furthermore, the recurrence relations satisfied by the orthogonal
polynomials provide a link with local equations of motion in the other
two approaches. 

The equivalence between Migdal and deconstruction correlators for
finite  $N$ gives a nice explicit realization of the so-called UV/IR
relation between the energy scale and the radial 
position in holographic models.
Indeed, in Migdal's approach the successive Pad\'e 
approximants allow an extrapolation of the UV result to lower and lower
energies.  
This corresponds in deconstruction to the addition of new
sites and links, which in turn generate an extra dimension.
Another interesting common feature is that a discrete spectrum
is obtained thanks to a violation of quark--hadron duality, which is
introduced by hand, either by keeping a constant $\tilde{t}=N^2
p^2/\mu^2$ in the Migdal limit or by cutting off the space with the IR
brane. Locality in $N$ implies that this violation is $\sim
1/\tilde{t}^N$, and exponentially suppressed in the continuum.

We have considered the large-$N$ Migdal limit, in which 
$N^2 p^2/\mu^2$ is constant. 
This  forced us to restrict the input functions to the
ones that have a conformal form. On the other hand, conformality in
the UV is related to asymptotically-$\ads_5$ spaces. The non-trivial fact
that we have explored here is that Migdal's approximation extrapolates
this conformal/AdS character all the way down from the UV to the IR, up
to an abrupt IR cutoff/brane.\footnote{We have also seen that a
  similar extrapolation is at work in other conformally flat spaces in
  the massless case.} This is 
related to the particular limit we are considering. It would be
interesting to investigate non-conformal input
functions, which arise at higher orders of perturbation theory.
This would require a different Migdal limit,
 and it could be speculated that a softer IR cutoff would be generated
 (possibly involving an infinite extra dimension, as in~\cite{cr}).
  Conformality is also broken by power corrections  which, in this context, were discussed in~\cite{erkrlo}. 
It would  also be interesting to study the correspondence for higher-point
 correlators~\cite{higherpoint}.

The relation between Pad\'e approximation and holographic calculations in 5D or
deconstruction could be regarded as a mere mathematical curiosity. 
However, we expect it to have physical consequences as well. 
In fact,
the Migdal program relies on dispersion relations and is similar in
spirit to the SVZ sum rules, which 
have a solid theoretical basis.
The connection with 5D models  might shed some light on the unexpected success of AdS/QCD models \cite{adsqcd}.
Furthermore,
Pad\'e approximation is often employed as a
unitarization method to extrapolate the predictions of chiral
perturbation theory in QCD\cite{padeqcd} and in no-Higgs models of
electroweak breaking \cite{padeew} to higher energies. This
approach is complementary to Migdal's, for it goes from low to high
energies rather than the other way round.\footnote{In~\cite{peris}, it has
been pointed out that, because the exact (large-$N_c$) two-point
function is a Stieltjes 
function, if the chiral contributions to all orders
were known, then the large-$N$ limit of the corresponding Pad\'e
approximants would exactly reproduce the full two-point function, at
arbitrary 
momenta. In particular, this puts restrictions on the allowed
chiral coefficients, directly related to the ones which can be derived
from 
dispersion relations~\cite{disprel}.} On the other hand, alternative
unitarization procedures using the 
notion of extra dimensions have been introduced more recently: the
so-called higgsless electroweak breaking \cite{higgsless}
and its deconstructed version \cite{dechiggsless}. Our results suggest
that these seemingly unrelated approaches could be equivalent,
although to prove it we should study the Pad\'e approximants
with a low-energy, rather than high-energy, input. 

\section*{Acknowledgements}

We thank Ian~Low and Alex~Pomarol for useful discussions.
A.F.\ 
is partially supported by the European Community Contract
MRTN-CT-2004-503369 for the years 2004--2008 and by the MEiN grant 1
P03B 099 29 for the years 2005--2007. 
M.P.V.\  
is partially supported by MEC (FPA 2003-09298-C02-01) and by
Junta de Andaluc{\'\i}a (FQM 101 and FQM 437).

\renewcommand{\thesection}{Appendix \Alph{section}} 
\renewcommand{\theequation}{\Alph{section}.\arabic{equation}} 
\setcounter{section}{0} 
\setcounter{equation}{0}

\section{Derivation of the boundary effective action in deconstruction}
\label{a.ba}

We derive here the holographic formula for the two-point correlation function of the ``boundary'' fields in deconstruction. 
We work in the tilted deconstruction framework; the minimal deconstruction result can be obtained by setting $\alpha_j = 0$.

The tilted deconstruction action can be rewritten as 
\beq
\label{e.atda} 
S = {1 \over 2}\int {d^4 p \over (2 \pi)^4}   
 \left \{\sum_{j,k=0}^N A_\mu^j D_{\mu \nu}^{jk} A_\mu^k 
 \right \}  \,,
\eeq 
where the kinetic operator is defined as  
\bea & \ds 
D_{\mu\nu}^{jk} = (-p^2 \eta_{\mu\nu} + p_\mu p_\nu) \left ( 
{1\over g_j^2} \delta_{j,k} 
+ {\alpha_j \over g_j^2} \delta_{j-1,k} 
+ {\alpha_{j+1} \over g_{j+1}^2} \delta_{j+1,k}  \right )
\nl \ds 
+ \eta_{\mu \nu} \left (
(v_j^2 + v_{j+1}^2)\delta_{j,k} 
- v_j^2 \delta_{j-1,k} 
- v_{j+1}^2 \delta_{j+1,k} \right ) \, .
\eea
For $D_{\mu\nu}^{00}$, $v_0 \equiv 0$ is understood. 
Our objective is to obtain an effective action for the boundary field $A_\mu^0$ after integrating out all resonances $A_\mu^j$ with $j \geq 1$. 
At tree-level, this can be achieved  by solving the equations of motion for the resonances with the boundary background field switched on,
\beq
\sum_{k = 0}^N D_{\mu \nu}^{jk} A_\nu^k  = 0  \qquad j \geq 1  \, ,
\eeq
and inserting the solution back into the action \erefn{atda}.
The solution can be written as  
\beq
A_\mu^j = \left (\eta_{\mu\nu} - {p_\mu p_\nu \over p^2}\right ) 
{F_N^j(p^2)  \over F_N^0(p^2)} +  {p_\mu p_\nu \over p^2}{F_N^j(0)  \over F_N^0(0)} \, ,
\eeq
where $F_N^j(p^2)$ solves the recurrence relation in $j$, 
\beq
\left ( v_{j+1}^2 + v_j^2 -  {p^2 \over g_j^2} \right ) F_{N}^j  
-  \left ( v_j^2 + {p^2 \alpha_j \over g_j^2} \right )F_N^{j-1}  
- \left (v_{j+1}^2 + {p^2 \alpha_{j+1} \over g_{j+1}^2} \right ) F_N^{j+1} = 0 \, ,
\eeq
subject to appropriate boundary conditions: $F_N^N =0$ in the Dirichlet case and 
$F_N^{N+1}=F_N^N $ in the Neumann case.  
Inserting the solution back into the action we obtain 
\beq
S_{\rm eff} = {1 \over 2} \int {d^4 p \over (2 \pi)^4} v_1^2  \left \{  
\left (-\eta_{\mu\nu} + {p_\mu p_\nu \over p^2}\right ) A_\mu^0 \Pi(p^2) A_\nu^0 
+  {p_\mu p_\nu \over p^2} A_\mu^0 A_\nu^0 {F_N^1(0) - F_N^0(0) \over  F_N^0(0)} 
\right \} \, .
\eeq
The second term vanishes in the Neumann case, while in the Dirichlet case it is cancelled by tree-level exchange of a massless physical Goldstone boson.
Finally, the polarization operator is given by
\beq
\label{e.tpodg} 
\Pi(p^2) = 
{F_N^1(p^2)  \over F_N^0(p^2)} \left ( 1 + {\alpha_1 \over g_1^2 v_1^2} p^2 \right ) 
+ {1 \over g_0^2 v_1^2} p^2 - 1 \, .
\eeq


\end{document}